# Hyperthermia HeLa cell treatment with silica coated manganese oxide nanoparticles

*A Villanueva[1], P de la Presa[2,3], J M Alonso[2,4], T Rueda[2], A Martínez[2], P Crespo[2,3], M P Morales[4], M A Gonzalez-Fernandez[4], J Valdés[2] and G Rivero[2,3,5]*

[1]Departamento de Biología. Universidad Autónoma de Madrid. Cantoblanco. 28049 Madrid Spain.

[2]Instituto de Magnetismo Aplicado (ADIF-UCM-CSIC), P.O. Box 155, Las Rozas, Madrid 28230, Spain

[3]Departamento de Física de Materiales, UCM, Ciudad Universitaria, 28040 Madrid, Spain.

[4]Instituto de Ciencia de Materiales de Madrid, CSIC, Madrid, Spain

Title running head: Hyperthermia HeLa cell treatment

**Abstract.** The effect of a high frequency alternating magnetic field on HeLa tumour cells incubated with ferromagnetic nanoparticles of manganese oxide perovskite $La_{0.56}(SrCa)_{0.22}MnO_3$ have been studied. The particles were subjected to a size selection process and coated with silica to improve their biocompatibility. The control assays made with HeLa tumour cells showed that cell survival and growth rate were not affected by the particle internalization in cells, or by the electromagnetic field on cells without nanoparticles.

However, the application of an alternating electromagnetic field to cells incubated with this silica coated manganese oxide induced a significant cellular damage that finally lead to cell death by an apoptotic mechanism. Cell death is triggered even thought the temperature increase in the cell culture during the hyperthermia treatment is lower than 0.5 ºC.

---

[5] to whom correspondence should be addressed, pdelapresa@adif.es; grivero@adif.es

# 1. Introduction

The treatment of tumours by hyperthermia is based on the different behaviour of normal and tumour cells versus temperature; generally, normal cells show better resistance to temperature than the tumour ones.[1,2,3,4,5] By taking advantage of this difference in the thermal resistance, it is possible to kill tumour cells selectively. In the last few years, magnetic nanoparticles (NPs) have attracted much attention for this medical application since they can be used as heat sources for magnetic hyperthermia. Under the influence of a high-frequency alternating magnetic field, they generate heat through hysteresis losses, induced eddy currents, Neel and Brown relaxation processes.[6,7,8] In this sense, numerous research works concerning hyperthermia have been centered in the study of new NPs with improved therapeutic efficiency. Superparamagnetic iron oxide is the most common material tested up to day due to its high biocompatibility, low synthesis cost, enhanced specific loss power and easy functionalization.[9,10,11] However, in spite of these advantages, is not possible to control the temperature distribution which depends on a large number of particle parameters such as size, concentration in tissue, conditions of the external applied field, and the length of the treatment,[12,13] apart of the tumor environment.

To avoid the obstacle of temperature controlling, new materials of tunable Curie temperature ($T_c$) are intensively investigated.[14,15,16,17] The new materials must satisfy strict conditions such as to be biocompatible (non toxic), stable in aqueous solution, possess high thermal efficiency as heating elements and have a high capacity of accumulation inside tumor cells so that when applying the alternating magnetic field (AMF) the increment of temperature induces the cellular death.[18,19] Magnetic particles with tuneable Tc will prevent that the temperature of the whole tumour or the hottest spot around the particle raise over the Curie Temperature, avoiding the use of any local temperature control system.

Recent reports claim the need to investigate new magnetic NPs with high magnetic moment as Fe or FeCo NPs or, on the contrary, the Fe oxide particle concentration in tumour must be higher in order to obtain larger specific absorption rate values.[20] However, it is still unclear whether it is necessary to increase the temperature up to 42 – 44 ºC in the whole tumour to induce tumour damage or it is enough to raise the temperature locally in cells to induce apoptotic tumour death. In the last case, materials with lower magnetisation could be also useful for hyperthermia treatments. The main objective of this work on intracellular hyperthermia is to enlighten about these questions. For this purpose, we study here the effect of applying an AMF after manganese perovskite incubation in HeLa cells and the analysis of the induced cellular damage or cell death mechanism.

The NPs are manganese oxides perovskite $La_{1-x}(SrCa)_xMnO_3$ with a Curie temperature that, depending on the cation ratio, can range from 300 K to 350 K and have large magnetization values of about 30 – 35 emu/g. The particles are obtained by ball milling method, subsequently subjected to a size selection process and coated with silica to achieve stability in water at pH 7 and high concentrations.

## 2. Experimental and methods

### 2.1 *Nanoparticle synthesis and characterization*

Powdered samples were synthesized by the ceramic method. Stoichiometric amounts of $La_2O_3$, $CaCO_3$, $SrCO_3$, and $MnO_2$ were homogenized and milled in an agate mill and then fired at 1400 ºC for 100 h in air to obtain $La_{0.56}(CaSr)_{0.22}MnO_3$ perovskite. The samples were finally quenched to room temperature. Cationic composition, as determined by atomic absorption, induced coupled plasma spectroscopy and electron probe microanalysis, is in agreement with the nominal one. The samples were undergone to mechanical milling for 1 h, in order to reduce the size.

Particles were dispersed in ethanol and heated over the $T_c$ in order to disaggregate them and to select the smaller ones. Briefly, the particles (800 mg) were dispersed in ethanol (200 ml) and the solution was put in an ultrasonic bath at 50 °C during 2 h. Then, the suspension was taken out the bath and left at 60 °C with reflux during 1 day. The solution was let settled at room temperature during one day, the largest particles tend to aggregate and they are collected with the help of a magnet. The particles, which still remain dispersed in the solution, are the smallest ones. The final yield is about 200 mg.

The particles were coated with silica following the Stöber method.[21] A thin silica layer was deposited on their surface at a constant temperature of 20 °C. The nanoparticles (20 mg) were added to a solution of 100 ml of 2-propanol that contained distilled water (5 ml) and ammonium hydroxide (1 ml). The solution was maintained in an ultrasonic bath for 1 h. Then, tetraethoxysilane (TEOS) (0.3 ml) was added to the solution and sonicated 10 min. This process was repeated twice. Finally, the solution was left in the ultrasonic bath for 5 h. The solution was filtered, and the nanoparticles were washed with 2-propanol and dried at 20 °C for one day. Then, they were dispersed again in distilled water.

Particle size was determined from transmission electron microscopy (TEM) micrographs in a 200 kV JEOL-2000 FXII microscope. For the observation of the sample in the microscope, a drop of the suspension was placed onto a copper grid covered by a carbon film and was allowed to evaporate. The mean particle size, d, was obtained from digitalized TEM images by counting more than 100 particles. Because the particles have irregular shapes, the maximum Feret's diameter, i.e., the maximum perpendicular distance between parallel lines which are tangent to the perimeter at opposite sides, is used to compute the size.

Dynamic light scattering (DLS) measurements have been performed in a ZETASIZER NANO-ZS device (Malvern Instruments Ltd, UK) to determine the hydrodynamic size of the silica coated manganese oxide perovskites (PER) in a colloidal suspension. The samples have been previously diluted in water in order to avoid multiple diffusion effects that reduce the hydrodynamic radius and increase the signal-noise ratio. The zeta potential was measured as a function of pH at 25 ºC, using $10^{-2}$ M $KNO_3$ as electrolyte and $HNO_3$ and $KOH$ to vary the pH of the suspensions.

Nanoparticles have also been magnetically characterized by mean of a Quantum Design SQUID magnetometer. The magnetic characterization consists in magnetization curves as function of temperature from 5 K to 350 K at 1000 Oe external applied field.

*2.2. Biological assays*

*2.2.1 Cell culture*

HeLa (human cervical adenocarcinoma) cells were grown as monolayers in Dulbecco's modified Eagle's medium (DMEM), supplemented with antibiotics and 10% fetal calf serum (all from Gibco, Paisley, UK). Cells were grown at 37°C in a humidified atmosphere containing 5% carbon dioxide. Cells in log growing phase were used for all experiments. For treatments, an appropriate number of cells were plated into 35-mm tissue culture plates with or without cover-slides and incubated to allow attachment. For cytotoxicity studies, cells were grown in 24-well tissue culture plates.

*2.2.2 PER internalization*

In order to analyse the internalization of PER, HeLa cells were grown on coverslips and incubated for 3 h with 0.5 mg/ml PER in DMEM supplemented with 10% fetal calf serum. After incubation, the containing medium was removed; the cells were washed three times with phosphate buffered saline (PBS) and observed immediately under bright light microscopy.

*2.2.3 Intracellular localization of PER nanoparticles*

To determine the intracellular localization of PER nanoparticles, the endocytic compartments of the HeLa cells were labelled with the fluoroprobe LysoTracker Red DND-99 (50 nM, Molecular Probes, Eugene, Oregon) in the culture medium at 37 ºC for 30 min. Previously, the cells were incubated with PER nanoparticles (0.5 mg/ml) for 3 h. After labelling, the coverslips were washed with PBS and were observed in a microscope under bright light illumination or fluorescence (green excitation filter) to detect the internalized PER nanoparticles and the emission of LysoTracker, respectively.

*2.2.4 Cytotoxicity*

The viability of HeLa cells incubated with PER nanoparticles was determined using the standard methyl thiazol tetrazolium bromide (MTT) assay (Sigma, St Louis, USA). After 24 h incubation with nanoparticles, MTT was added to each well (the final concentration of MTT in medium was 50 μg/ml) for 3 h at 37 ºC. The formazan formed in the cells was dissolved adding 0.5 ml of DMSO in each dish, and the optical density was evaluated at 570 nm in a microplate reader (Tecan Spectra Fluor spectrophotometer). Cell survival was expressed as the percentage of absorption of treated cells in comparison with that of control cells (not incubated with PER nanoparticles). The results obtained are the mean value and standard deviation (SD) from three experiments.

*2.2.5 Alternating magnetic field (AMF)*

HeLa cells were incubated with 0.5 mg/ml PER for 3 h. After incubation, cells were washed 3 times with PBS and then exposed for 30 min to an AMF (f = 100 KHz and H = 15 mT). Control plate cell PER-free was exposed to magnetic field. The temperature of the cultures was controlled throughout the complete experiment by an infrared thermometer. The temperature of the cell culture at the beginning of the treatment was 37 ºC. Different methodological protocols were carried out 24 h after application of AMF.

*2.2.6 Apoptotic studies*

Apoptotic morphological changes were analyzed for assessing cell death following PER internalization and application of AMF. Cells were fixed with ice-cold methanol (5 min), air-dried and stained with Hoescht-33258 (10 μg/ml in distilled water, 5 min) for chromatin visualization. After washing and air-drying, preparations were mounted in DPX and observed under fluorescence microscopy. Identification of apoptotic nuclei by staining with Hoechst-33258 was based on key features of apoptosis, namely condensation and fragmentation of nuclear material. The proportion of apoptotic nuclei was determined by counting in a fluorescent microscope with apoptosis defined as the presence of two or more condensed bodies per nucleus.

The effect of PER + AMF on cell detachment in adherent HeLa cells was also investigated. Detached cells were collected and centrifuged at 1200 rpm (radius: 7 cm) for 5 min, washed in PBS, centrifuged again, and fixed with 0.25 ml of 70% cold ethanol. Cell suspensions were stained with Hoechst-33258 at a final concentration of 20 μg/ml for 5 min. A drop of cellular suspension was mounted between a microscope slide and a cover slide and visualized by epifluorescence microscopy. Apoptotic cells were identified using morphological criteria (chromatin condensation and fragmentation).

Detached and attached cells were counted in a hemocytometer. We calculated the percentage of detachment by dividing the number of detached cells by the total number of cells. The percentage of apoptotic cells was calculated as the ratio of apoptotic detached cells divided by the total number of cells in three independent experiments at several magnetic fields with the fluorescent microscope. For each experiment, 500 cells per treatment were counted.

*2.2.7 Microscopy*

Microscopic observation and photographs were performed in an optical Olympus BX61 microscope equipped with ultraviolet and green exciting filter sets for fluorescence microscopy and an Olympus DP50 digital camera; all photographs were taken using Photoshop CS software (Adobe Systems, San Jose, USA).

## 3. Results and Discussion

### *3.1. Magnetic nanoparticles and its characterisation*

The particles obtained by ball milling method form agglomerates due to the dipolar magnetic interaction and the lack of surfactants. The agglomerates have a large size distribution with sizes greater than 1 μm (figure 1A).

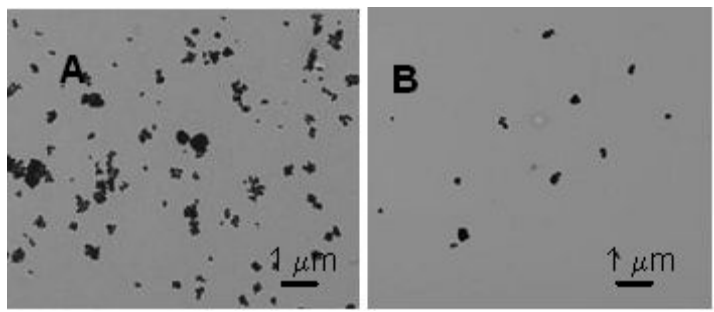

**Figure 1**. TEM images of PER nanoparticles: A) After ball milling; B) After size selection by ultrasonic treatment above Curie temperature.

In spite of the lack of homogeneity in size, the ceramic synthesis method has the advantage of allowing the formation of materials with complex cationic compositions like this $La_{0.56}(CaSr)_{0.22}MnO_3$. The method assures chemical homogeneity of the sample, which is of relevance to control Tc.[22] After Tc determination, particle size is selected by means of their own magnetic characteristics. Since dipolar magnetic interactions disappear over Tc, the particles disaggregate easily with increasing temperature. When particles are dispersed in ethanol, sonicated and heated over Tc, aggregates are broken, producing more isolated NPs of smaller sizes (figures 1B). Particle size distribution analysis obtained from the TEM pictures leads to an average size of 150 nm and a standard deviation of 30% (figure 2).

Finally, magnetic nanoparticles were coated with a $SiO_2$ shell rather homogeneous. Silica is coating single particles as well as aggregates as shown in figure 3. The average thickness of the silica coating (5-10 nm) was extracted directly from the TEM images.

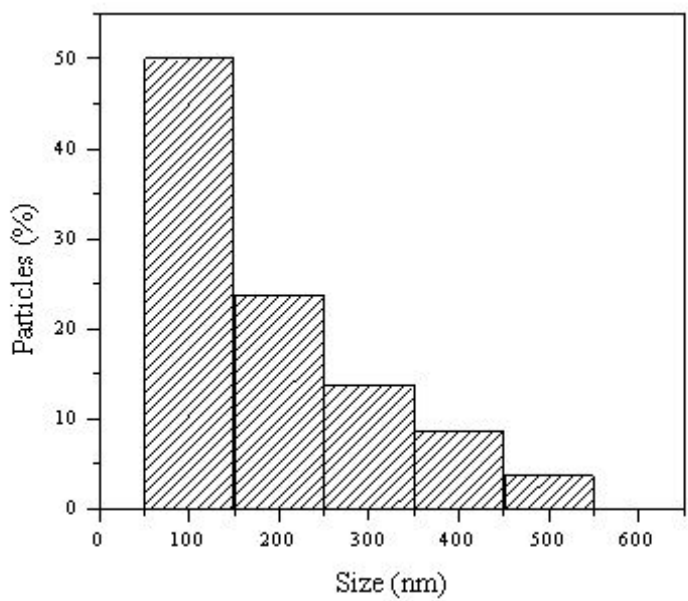

**Figure 2.** PER nanoparticles size distribution calculated from TEM data.

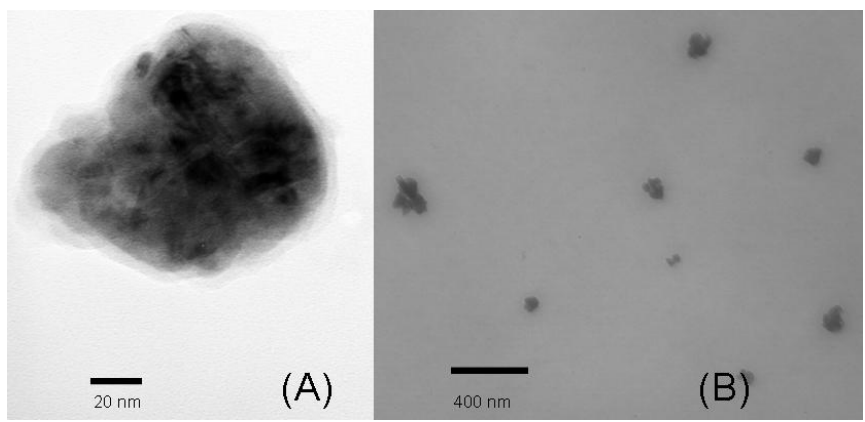

**Figure 3.** TEM images of silica coated single PER nanoparticles at high (A) and low (B) magnification.

The silica coating at the manganese oxide surfaces stabilizes the particles in aqueous suspension under physiological conditions (pH and salinity). The mean hydrodynamic size, which is the effective aggregate size of the particles in solution, is about 300 nm with a standard deviation σ = 0.45 (figure 4). This is at the limit size to fit the requirements for intravenous injection and therefore for in vivo biomedical applications.[23],[24]

The presence of silica on the surface of the perovskites results in a shift of the isoelectric point towards lower pH values, i.e. from pH 8 to near pH = 3. The variation of the zeta potential as a function of pH for naked (open circles) and coated nanoparticles (fill circles) is shown in figure 5. The surface potential is obviously different for uncoated and coated particles. At pH 7, for instance, the surface potential is -20 mV for the silica-coated sample, and +2.5 mV for the uncoated ones, confirming that silica coating stabilize PER in water by both, steric and electrostatic repulsion. These colloidal suspensions are stable in water at concentrations as high as 5 mg/ml.

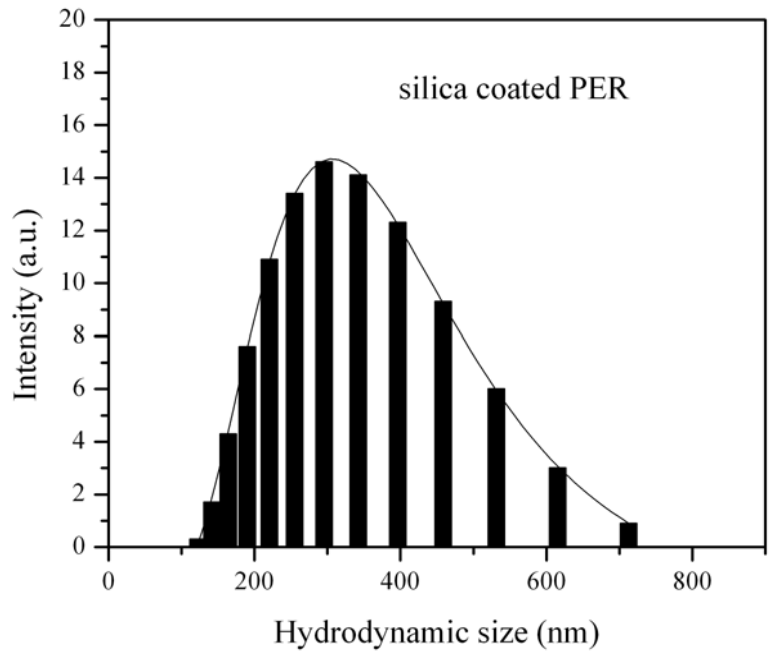

**Figure 4.** Hydrodynamic size distribution for PER nanoparticles in water at pH 7.

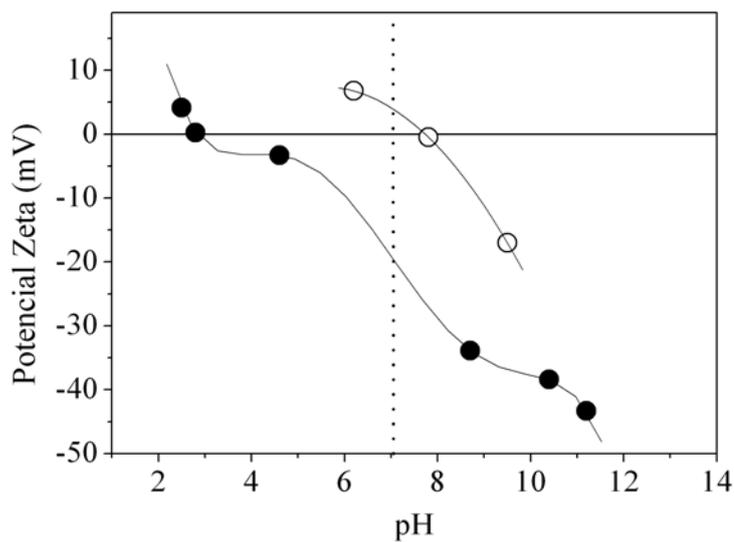

**Figure 5.** Z potential versus pH for PER nanoparticles as prepared (open circles) and silica coated nanoparticles (full circles).

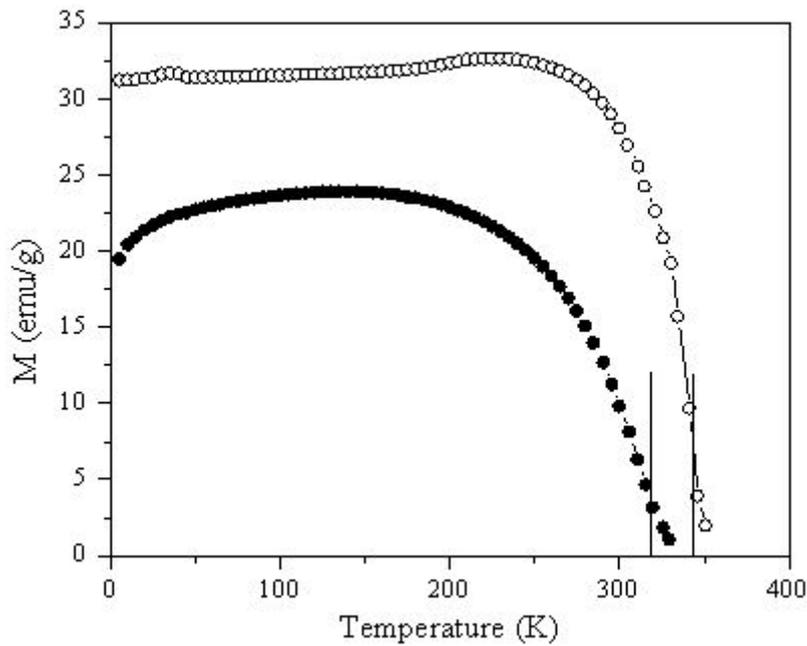

**Figure 6.** Magnetization curves versus temperature for PER nanoparticles as prepared (open circles) and silica coated nanoparticles (full circles).

Magnetic properties, in terms of saturation magnetisation and Curie temperature, are not significantly affected by the size selection process. Mean particle size is above 100 nm which means that surface effect are diminished, magnetisation values are preserved and Curie temperature is kept at 68 ºC. Below at certain particle size, surface effects begin to be important and should be considered to understand the magnetic behaviour of the particles.[25] However, magnetic properties are strongly affected by the coating: i) total magnetization is reduced by the presence of diamagnetic silica and ii) Tc decreases from 68 ºC to 44 ºC as observed in Figure 6 where magnetization curves for both samples at room temperature are presented.

At low temperature, the magnetization at 1 kOe decreases from 31 to 21 emu/g (about 32%) for the uncoated and the coated PER samples, respectively. Considering PER samples consisting of 100 nm spherical manganese oxide cores and a silica shell of 10 nm thickness, as a rough approximation, the total volume of silica represents about 35 % of the whole PER particle. Therefore, the large amount of diamagnetic silica can account for the decrease in the total magnetization values. This reduction in magnetisation is expected to reduce the heating efficiency of the material as it has been observed for iron oxide nanoparticles.[12]

On the other hand, Tc reduction is better related to the interaction of silica with atoms at manganese oxide surface. The presence of silica reduces Tc by only 7%, from 341 K (68 ºC) to 317 K (44 ºC) (Fig. 6). However, considering the narrow temperature range at which MPH treatments are performed, the decrease from 68 to 44 ºC is significant.

Therefore, a compromise should be achieved about the silica coating which should be thick enough to assure water stability at high nanoparticle concentrations while keeping the magnetisation as high as possible to preserve the heating efficiency. Moreover, magnetic nanoparticle silica coating affects strongly PER Curie temperature and can be used for tuning the switching temperature of the material in self-control hyperthermia treatments.

*3.2. PER internalization and subcellular localization*

HeLa cells incubated for 3 h with 0.5 mg/ml PER and visualized by optical microscopy, showed an intracellular punctual distribution consisting in black cytoplasmic spots of different sizes. They are not visualized in control cells without PER incubation (Fig. 7A and 7D respectively). PER nanoparticles can been detected directly inside the cells under bright light microscopy without being processed to avoid potential artifacts of cell fixation. It is not necessary to use complementary techniques (TEM or fluorescent-labelled nanoparticles) for their visualization, as it happens for most of the studied magnetic nanoparticles.[26,27]

Several studies have evidenced that the incorporation entrance mechanism of NPs into the cell is endocytosis or phagocytosis (in cells of the reticulated endothelial system). LysoTracker Red DND-99, a specific probe for acidic compartments (endosomes), was used to confirm this point. As shown in Figure 7 C, a substantial fraction of the red fluorescence from the LysoTracker dye co-localizes with the black spots of internalized PER nanoparticles. A similar subcellular distribution has been described for other nanoparticles, and seems to indicate that the mechanism of nanoparticles entrance into the cell is

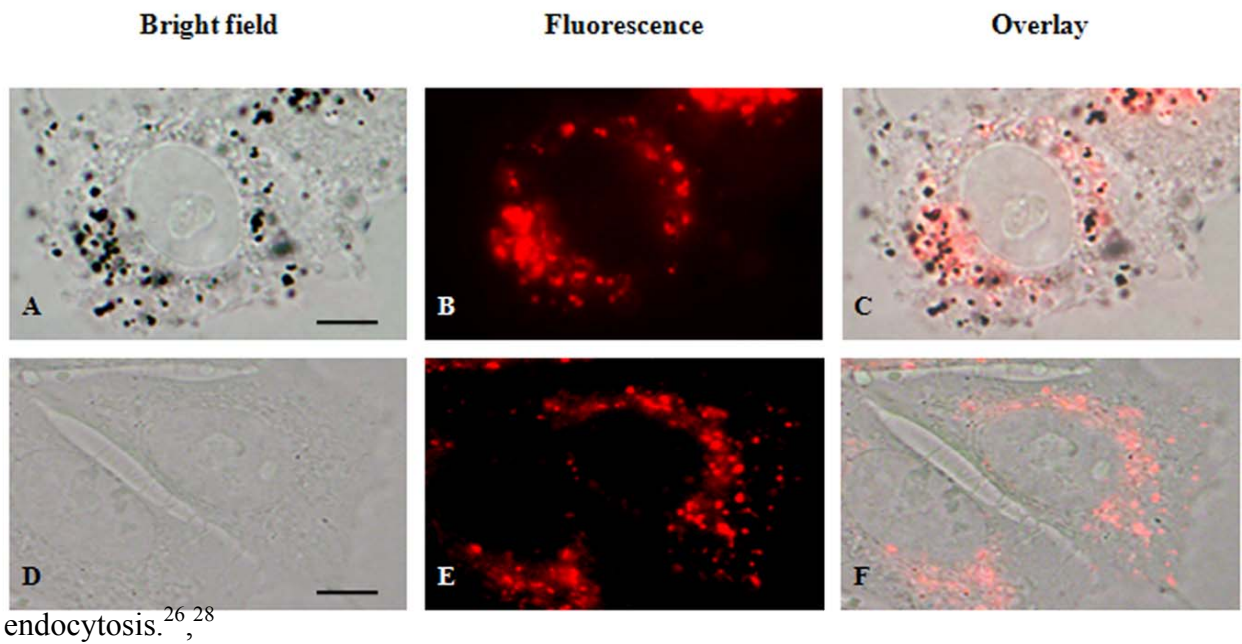

endocytosis.[26,28]

**Figure 7.** Localization of PER nanoparticles and accumulation of LysoTracker Red in endocytic compartments. **A**: HeLa cells incubated with 0.5 mg/ml PER for 3 h in bright-field microscopy. **B**: Localization of LysoTracker in the same cell. **C**: Merged images. **D-F**: Control HeLa cells visualized with the same assay. Scale bar: 5 $\mu$m.

Biocompatibility of PER (0.5 mg/ml) was evaluated by means of the standard MTT assay. The cytotoxic analysis after 3 h of incubation of HeLa cells with PER showed that the viability of cell culture was not significantly affected by the presence of PER after 24 h post-incubation (98.7 ± 2.1 % viability in relation to the control sample). The toxicity seems to depend on the surface, size and type of nanoparticles.[29]

Hoechst-33258 staining revealed the absence of morphological alterations after incubation with PER. HeLa cells interphase has an oval nucleus containing several nucleoli. PER incubated cells and stained with Hoechst-33258 showed a nuclear morphology very similar to controls (see figure 8A and 8B).

### *3.3. Alternating magnetic field treatment*

The nuclear morphology was not affected for the corresponding controls: HeLa cells without any treatment, only PER incubation without AMF exposition, AMF without PER pre-incubation. However, the PER + AMF treatment provoked deep morphological alterations 24 h after the combined treatment, which corresponds to a cell death by an apoptotic process. The achieved temperature of the cell culture during the PER + AMF treatment was lower than 37.5 ºC. As it can be seen in Figure 8C, a significant number of cells (19.6 ± 7.8 %) showed round shape, shrunk and a typical chromatin fragmentation visualized by Hoechst-33258 staining. Morphological alterations by PER + AMF exposure was accompanied by cell detachment. A significant attached HeLa cells (17.8 ± 8.3%), lose adhesion to the substrate and, therefore, appeared floating in the culture medium 24 h after treatment. The fraction of detached cells were apoptotic (>98%). It is interesting to note that cells with apoptotic morphology still showed PER nanoparticles inside (see figure 8D-G). Shrinkage of cell volume, condensation and fragmentation of chromatin are morphological characteristics of apoptosis.[30,31,32] Similarly, loss of adhesion to the substrate called "anoikis" is clearly related to apoptosis.[33]

It must be highlighted that the morphological alteration of the cells under PER + AMF treatment took place even though the temperature increase of the cell culture was lower than 0.5 ºC.

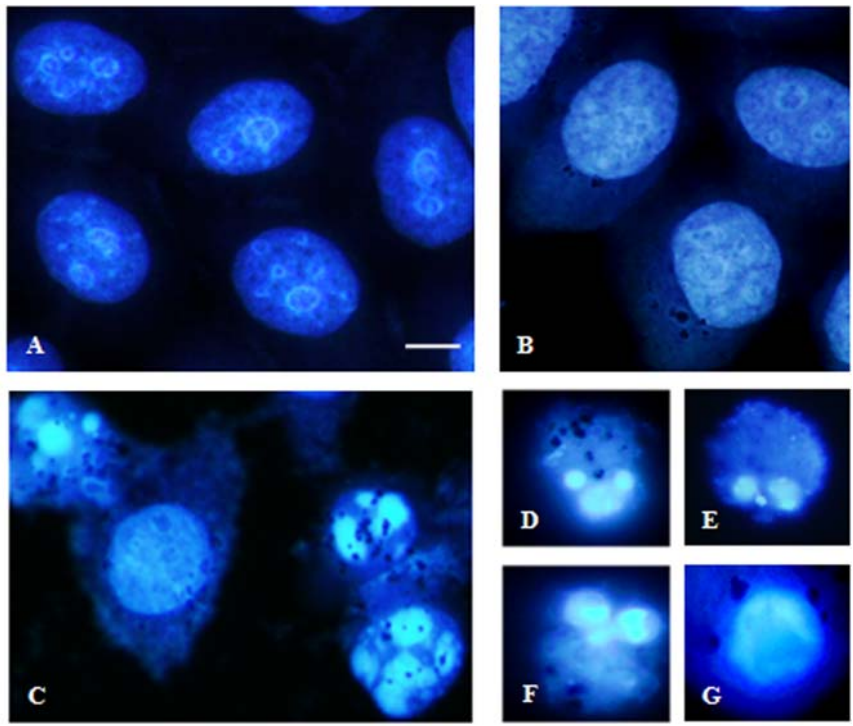

**Figure 8. A**: Interphase HeLa control cells stained with Hoechst-33258. **B**: Interphase cells treated only with PER and stained with Hoechst-33258. **C**: Morphological changes of attached cells HeLa cells induced by PER + AMF 24 h after treatment. **D-G**: Detached cells after combined treatment showing apoptotic nucleus. Scale bar 5 μm.

In summary, our results provide important information about the use of PER nanoparticles for cancer hyperthermia treatments. These nanoparticles are biocompatible and are capable of inactivating tumor cells in culture, after applying an alternating magnetic field. The scarce previous work on perovskites, for the same purpose,[16] do not include biological assays.

Further experiments will be conducted to analyze in more depth the effectiveness of these treatments.

**4. Conclusions**

The effect of applying an alternating magnetic field to HeLa cells after incubation with manganese perovskite nanoparticles has been studied and the induced cellular damage or cell death mechanism have been analysed. Magnetic manganese oxide nanoparticles have been coated with a silica shell achieving water stability at high concentrations and biocompatibility, i.e. large cell survival after 24 h. The application of an alternating magnetic field of 15 mT and 100 KHz during 30 min produced cellular damages that finally lead to apoptotic cell death even though the temperature increase in the cell culture was lower than 0.5 ºC.

**Acknowledgments**: This work was supported by the Spanish Ministry of Science and Innovation through projects MAT2005-06119 and CONSOLIDER on Molecular Nanoscience CSD 2007-00010 and by the Comunidad de Madrid under project Nanomagnet S-0505/MAT/0194. PdlP acknowledges support from the Spanish Ministry of Education and Science through the Ramon y Cajal program.